# Pupil plane wavefront sensing for extended and 3D sources

Roberto Ragazzoni[ab] Valentina Viotto[ab], Elisa Portaluri[ab], Maria Bergomi[ab], Davide Greggio[ab] Simone Di Filippo[bc], Kalyan Radhakrishnan[ab], Gabriele Umbriaco[c], Marco Dima[ab], Demetrio Magrin[ab], Jacopo Farinato[ab], Luca Marafatto[ab], Carmelo Arcidiacono[ab], Federico Biondi[ab]

[a]INAF – Osservatorio Astronomico di Padova, Vicolo dell'Osservatorio 5, 35122, Padova, Italy
[b]ADONI – ADaptive Optics National laboratory in Italy
[c]Università degli Studi di Padova – Dipartimento di Fisica e Astronomia, Vicolo dell'Osservatorio 3, 35122, Padova Italy

## ABSTRACT

The basic outline of a pupil plane WaveFront Sensor is reviewed taking into account that the source to be sensed could be different from an unresolved source, i.e. it is extended, and that it could deploy also in a 3D fashion, enough to exceed the field's depth of the observing telescope. Under these conditions it is pointed out that the features of the reference are not invariant for different position on the pupil and it is shown that the INGOT WFS is the equivalent of the Pyramid for a Laser Guide Star. Under these conditions one can imagine to use a Dark WFS approach to improve the SNR of such a WFS, or to use a corrected upward beam in order to achieve a better use of the LGS photons with respect to an ideal Shack-Hartmann WFS.

## 1. INTRODUCTION

A WaveFront Sensor (WFS hereafter) is a device that is able to convert the collected light in a manner that is possible - within certain limitations- extract information such that one can reasonably estimate the wavefront shape of the incoming light. In its more widely used form, the core of such a WFS is made by some optomechanical device that is able to convert the incoming beam in a pattern whose light distribution reflects in some manner the shape of the incoming wavefront. In this respect the WFS "rearrange" the incoming photons and, once detected, the intensities can be used to retrieve the desired information. A Shack-Hartmann WFS subdivide the light at the pupil plane level and rearrange in the form of spots, at least when the source is an unresolved source, such that their position is proportional to the first derivative of the incoming wavefront. A class of WFSs that gained more and more attention in the recent decades is the so-called class of pupil-plane WFSs. This is also due to the demonstrated higher sensitivity with respect to the once ubiquitous Shack-Hartmann, of the Pyramid WFS [1], belonging to such a class. In this kind of device (see Fig.1) some optomechanical "stuff" is located close to the focal plane (or, better, in the focal volume) and a reimager of the pupil plane will exhibits one or more pupil images properly perturbed. It is to be pointed out that as soon as the perturbator has some optical power, the pupil reimaging is no longer perfect and, assuming the optical power is modest, these would not be anymore, strictly speaking, pupil plane WFS but one could name them "quasi pupil plane WFS". A non exhaustive list of pupil-plane or quasi pupil plane WFS is given in the following along with a short comment.

- Foucault knife (it only measure one the derivative of the wavefront in one axis and it is sensitive to scintillation in the pupil plane);
- Curvature "a la Roddier" (a reflecting membrane is located on the focal plane so that the reflected beam is affected in a manner that slightly out of pupil plane imaging is achieved. With fast detector allow for Phase Locked Loop tracking of the signal, that is proportional to the second derivative, or Laplacian, of the wavefront)
- Point diffraction or Smartt (this kind of WFS produce an illumination directly proportional to the departure of the wavefront itself from the ideal flat one, but its dynamic range is rather limited);

- Ronchi ruling (a grating is located in the focal plane and again the measurement is accomplished only along one axis so that, like in the Foucault knife one have to split the beam in two before the sensing along two channels);

- Pyramid (it simultaneously record pupils in a manner that resemble the Foucault knife but allowing for being robust to scintillation in the pupil plane and estimating the derivative of the wavefront in an efficient manner in both axis. A modulation on the pin of the pyramid allow for any desired dynamic range, at the expenses of sensitivity, that has been demonstrated to be significantly higher than the most widely used in astronomy non pupil plane WFS: the Shack-Hartmann);

- Flat pyramid (the four pupils superimpose themselves but the combination of the light still allow for an efficient retrieval of the first derivative of the wavefront, especially for small perturbations);

- Dark WFSensing (the rationale behind this variation is to remove as much as possible the light that is common within the pupil plane within the range of the accepted perturbation)

- Generalized differentiation (it exhibits a non linear behavior such to incorporate the high dynamic range for large perturbations and the high gain achieved from the pyramid when the perturbation is rather small)

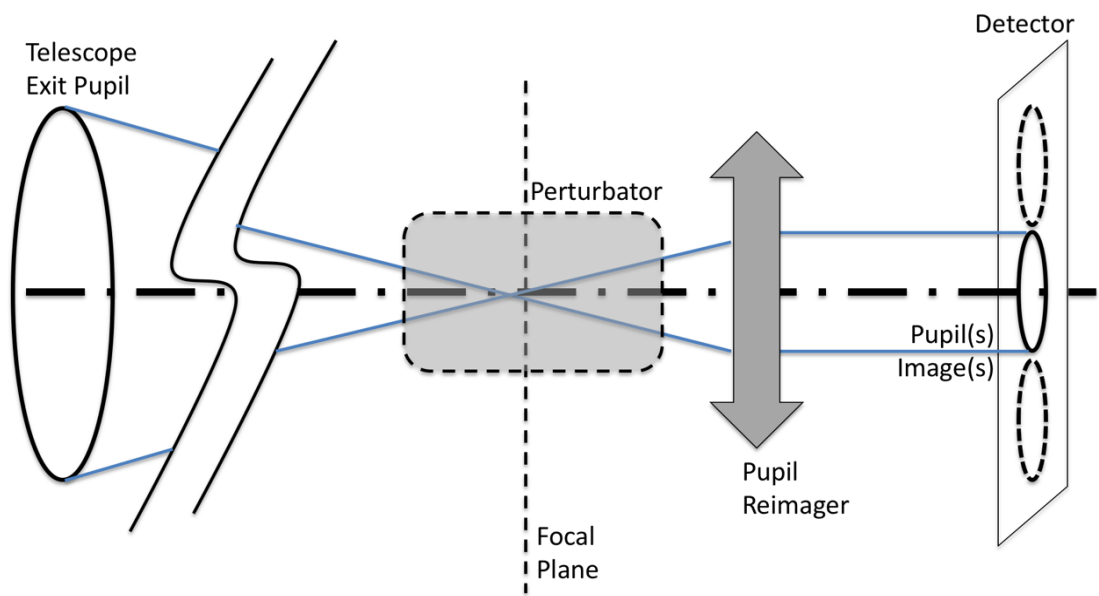

**Figure 1 A generic pupil-plane WFS. The incoming light beam is collected from the left to the right and in the volume close to where the reference source is focused a perturbator is being located. After that a pupil reimager made one or more pupil images on the proper plane whose illumination is somehow linked with the perturbation of the wavefront itself. If the perturbator has some optical power the nominally perfect pupil reimaging is lost and one can speak of quasi-pupil plane WFS.**

Further classes of pupil-plane WFS are the layer-oriented or the onduline one (to cope with multiple simultaneous unresolved references) or the z-invariant WFS that copes with a reference that is spread not only in the plane orthogonal to the optical axis but also along the same.

## 2. LASER GUIDE "STAR"

The Laser Guide Star, i.e. a reference beacon [3] produced by mesospheric excitation of the natural Sodium layer that envelope our planet at an altitude of approximately 92km, has several interesting features. First of all it has a finite distance, such that with basically any modern telescopes, the position where he will get in focus will be significantly displaced with respect to where heavenly bodies will get in focus. Furthermore, unless they are projected through the same, or a larger aperture from which it has to be sensed, through an AO upward correcting adaptive optics system (a very unlikely scenario) the reference is by no means unresolved and, finally, it deploy over a significant height, in other words, different portion of such a source will fall in focus positions significantly displace with respect to each other. However, there is a significant subdivision of the kind of WFS that one could conceive to cope with such kind of reference depending if the beacon is propagated from a projector that lie within the pupil of the telescope or outside. See Fig.2 for a clarification of the two different situations in modern two mirrors telescopes.

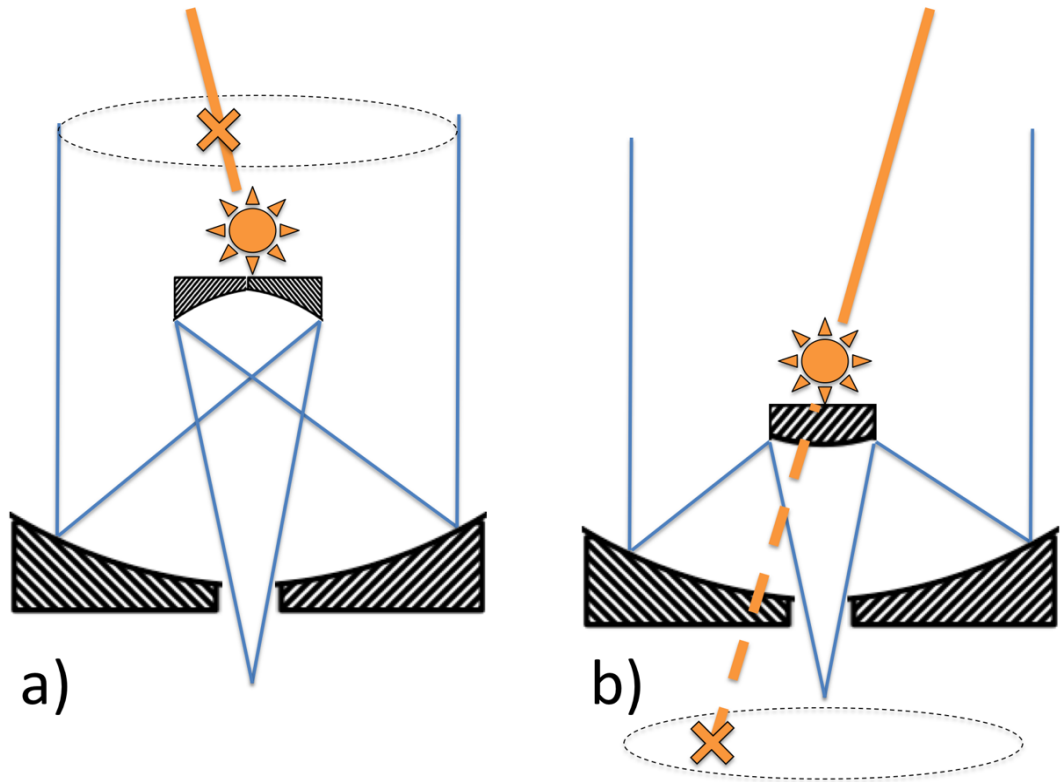

**Figure 2 In modern telescopes the entrance pupil is no longer located at the main primary mirror in order to properly reject that thermal InfraRed emission from the surrounding or the structure of the telescope itself. The pupil is in fact defined by the secondary mirror and in case of a Gregorian configuration (case a on the left) the entrance pupil is located somehow above the telescope along its optical aixs, while in a more conventional Cassegrain configuration (case b on the right) the entrance pupil is located somehow below the telescope tube. What does matter, for the purpose of defining the class of WFS that are to take care of an axially elongated beacon as the Sodium Laser Guide Starsm is the position where the propagation beam does intercept the entrance pupil plane.**

The two cases so far depicted produces situation in the volume close to the focal plane that are so different that completely different kind of perturbators are to be considered convenient. In the case where the LGS is propagated from within the entrance pupil the beam from a single point on the reference beacon is always "embedded" within the envelope of the other ones. This lead to solutions that belong to the class of the z-invariant WFS [3] with the noticeable, but surely suboptimal, exception, when the central obstruction is significant, of a "cascade" of pyramids that can lie within the shadow of the central obstruction [4] for most of the propagation time of the LGS. In the case where the LGSs are propagated from outside the entrance pupil the points where the LGS focus along its structured beacon stay outside the beam of the others portions. It is noticeable that when the LGS is fired just on the edge the loci of the focus point stay within the edge of the conical converging beam. This means that with a reasonable focal ratio the angle is rather steep, closer to the vertical rather than to the horizontal. While the actual angle is engineerable through the equivalent focal ratio where the LGS sensor is located (usually a dedicated port that can take advantage of the monochromatic nature of such a beacon) one should recall that a long focal ratio also translates into a physical large distance between the points where the LGS focuses. It should be reminded that during tracking the actual range of the LGS vary considerably in a manner that is now wildly because of the slow evolution of the altitude of the reference, but the peak to valley variations exceed by far the factor of 2 to 3, depending upon the ultimate choice in the minimum elevation of operation, reminding that at low altitude the equivalent airmass would unavoidably deteriorate any adaptive optics performance

## 3. STRETCHING THE PYRAMID FOR A GOOD REASON

Given the difference with respect to the reference source and how it deploys in the focal volume, one could be tempted to properly "stretch" a refractive pyramid ion order to cope with the faced situation. An obvious solution would be to populate the Scheimpflug plane where the various elements of the reference beacon goes in focus. A stretching of the pyramid would involve a splitting of the original four pupils to at least six faces and related pupils, and still giving up in this way the use of the (variable, and smoothed out in different portions of the pupil) structure along the beacon. Because of the high angles involved it is easy to see that with conventional refractive index material one is faced to both strong distortion of the bent light leading to deformed pupil images, and to a very inefficient refraction due to the high angle of attack and hence a very likely spurious reflection where most of the light can go. It is true that the monochromatic nature of the Sodium reference makes suitable an hypothetical proper coating, but such an option is just mention here and not further discussed.

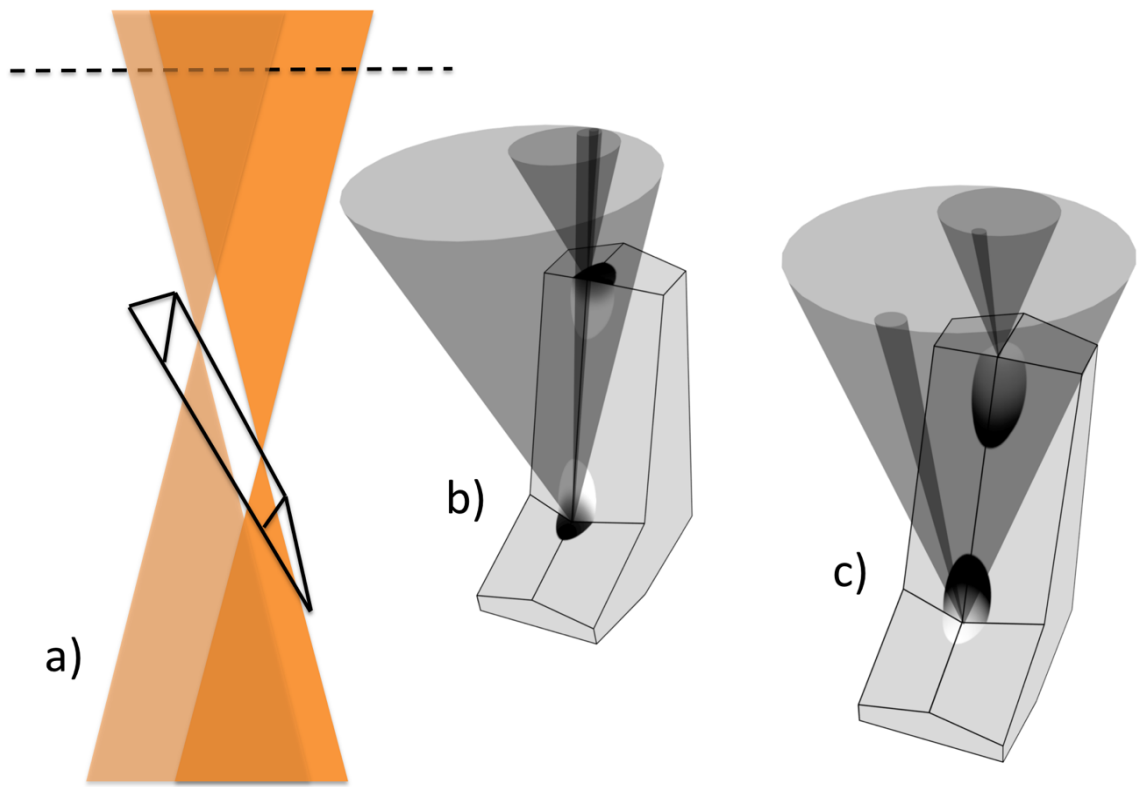

**Figure 3 The loci where different point of a Sodium layer beacon are reimaged stay on a steep line well distant from the dashed line where astrophysical objects get in focus (see the rightmost box a). An hard if not impossible refractive device would represent a sort of stretched pyramid. Because of these practicalities an ingot where the light is refracted or reflected and refracted, has been devised, in order to bent the beams to form six different pupil images on a common plane. It is noticeable that (central box b) the portion of the beam in the pupil region close to where the LGS is being fired "see" the ingot with such a grazing angle that most of the light is confined to the topmost and lowermost faces, makes in fact a four quadrant, pyramid like, sensing for the region of the aperture where the elongation effect is modest. In contrast, portion of the incoming beam at the largest distance from where the LGS is being fired (right most box c) will basically feed most of their light into the central elongated region,**

**hence pouring most of the photons for the solely determination of the derivative orthogonal to the apparent elongation.**

The main difficulty into coping with this approach is that the apparent width of the Sodium layer is continuously variable [5] and the optimal length of the ingot portion that is devoted to the solely sensing if the derivative of the wavefront orthogonal to the elongation is not a fixed figure. Some approaches has been devised out. One is to have a drum carrying several ingots such that the optimum one is chosen with a sort of revolving mechanism. Another option is to have a zoom optics such that the scale is adjusted in order to match a fixed ingot size. This last option, however, makes the pupil image changing accordingly unless a counter-zoom system is accomplished on the pupil reimager, noting that in this way the inter-distance between the pupils would change accordingly nevertheless. A sub-optimal solution where the ingot is deliberately tilted with respect to the incoming beam is hard, as the cosine law of changing the apparent projected size is unfavorable and, moreover, one would loose on the key reason to build the ingot, that is to have the loci of the focus of the beacon properly aligned with a sharp border between two faces of the prism splitting the light into the various pupil's images. Another interesting issue is represented by the problems faced at the acquisition. At the beginning of operations, in fact, it is unlikely the ingot is not even closely aligned. One can see that if an additional modulating movement is accomplished along the edge of the main edge of the ingot a signal is available on the pupil that is proportional and somehow orthogonal to the others, for any kind of misalignment with respect to the reimaged beacon. One should remind that such a modulation is not required during operations but only at the acquisition and eventually needed time to time during reacquisition for example after repointing. For all these reasons mentioned here one would like to simplify the ingot concept to what we called the 3-ingot (see also Fig.4) where one of the end of the sodium beacon is deliberately not taken into consideration [6,7,8]. With the proper choice of which end to take into account this represent less than the loss of half of the signal, simply choosing the sharper end.

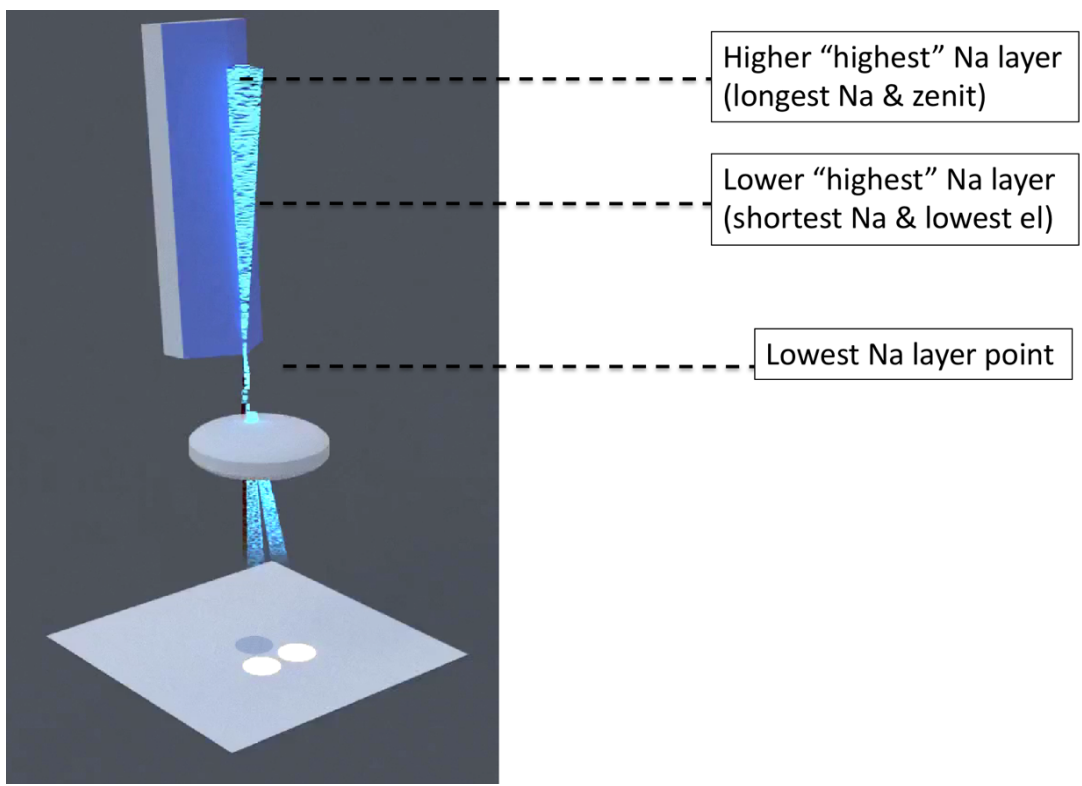

**Figure 4 A 3-ingot scheme in which a single roof prism is aligned with the reimaged loci of the sodium beacon. One of the end of the roof represent the interface with a direct pupil reimaging to sense the wavefront derivative along the apparent LGS elongation. The third pupil is reimaged directly while the other two are bent via a grazing reflection. This approach exhibits inherent no chromatism and because of the grazing incidence the requirement on the quality of the reflecting surfaces is rather loose. An optimal pupil reimager design will consider to have its**

**optical axis tilted with the respect to the main telescope optical axis as the direct pupil is on-axis while the other two are off-axis ones. During operations the lowest reimaged Sodium layer point is kept slightly off the lower edge of the roof while**

## 4. CONCLUSION

One of the main practical advantage of the pupil plane approach in both NGS and LGS wavefront sensing is represented by a very compact use of the detector. In fact even for ELT needs one is faced with the need of a detector barely larger than the minimum sampling required. For an 80x80 sampling a 256x256 detector is a doable approach, while any reasonable Shack-Hartmann solution would requires one or two orders of magnitude more pixels. WFS sensitivity, as soon as the LGS is being fired with an upward correction with a beam diameter larger than the sampled subapertures , is obtained, in contrast with other options. This is a rather interesting one because you can conveniently lower the laser power requirements, or allow for a much faster sampling in order to extend AO correction in the bluer side of the spectrum. A one meter diameter projector, with an upward AO assisted compensation, although not necessarily an easy task, can lead to a 4-fold advantage that could translates into four times less power or four times faster sampling. Assuming the upward beam is uncompensated, one could also take advantage of the dark WFS concept "obscuring" a strip of light along the edge in order to remove, to some extent, photons that would contribute to the photon shot noise and not necedssarily to the signal. This concept, however, is left at a speculative level at this point.